\documentclass[12pt]{article}
\def\be{\begin{equation}}					 
\def\ee{\end{equation}}
\def\ber{\begin{eqnarray}}
\def\eer{\end{eqnarray}}	
\begin{document}
\vspace*{1cm}
\begin{center}
{\Large \bf Mass Spectrum of
Three-Pion System \\[1ex] in Kaluza-Klein Picture}\\

\vspace{4mm}

{\large A.A. Arkhipov\\
{\it State Research Center ``Institute for High Energy Physics" \\
 142280 Protvino, Moscow Region, Russia}}\\
\end{center}

\vspace{4mm}
\begin{abstract}
{In this note we present additional arguments in favour of
Kaluza and Klein picture of the world. In fact, we show that
formula (\ref{KK3pi}) provided by Kaluza-Klein approach with the
fundamental scale early calculated \cite{1}  gives an excellent
description for the mass spectrum of three-pion system.} 
\end{abstract}

\section*{}

In our previous papers \cite{1,2} we have presented the arguments in
favour of that the Kaluza-Klein picture of the world has been been
observed in the experiments at very low energies where the
nucleon-nucleon dynamics has been studied. In particular we have
found that geniusly simple formula for KK excitations provided by
Kaluza-Klein approach gives an excellent description for the mass
spectrum of two-nucleon system. In article \cite{3} we have presented
additional arguments in favour of Kaluza and Klein picture of the
world. In fact, we have shown that simple formula provided by
Kaluza-Klein approach with the fundamental scale early calculated
\cite{1} gives an excellent description for the mass spectrum of
two-pion system. Surely, this was quite an event and,
certainly, this very nice fact encouraged us to continue the study of
the other hadronic systems in this respect. Taking this line we have
performed an analysis of experimental data on mass spectrum
of the resonance states of three-pion system and compared them with
calculated values provided by Kaluza-Klein scenario. In this note we
present the results of this analysis.
  
As in the previous cases let us build the Kaluza-Klein tower of KK
excitations for three-pion system by the formula 
\be
M_n^{\pi^1\pi^2\pi^3} = \sqrt{m_{\pi^1}^2+\frac{n^2}{R^2}} +
\sqrt{m_{\pi^2}^2+\frac{n^2}{R^2}} +
\sqrt{m_{\pi^3}^2+\frac{n^2}{R^2}},\quad
(n=1,2,3,\ldots),\label{KK3pi}
\ee
where $\pi^i(i=0,+,-)=\pi^0,\pi^+,\pi^-$ and $R$ is the same
fundamental scale calculated early from the analysis of
nucleon-nucleon dynamics at low energies \cite{1,2}
\be
\frac{1}{R} = 41.481\,\mbox{MeV}\quad \mbox{or}\quad
R=24.1\,GeV^{-1}=4.75\,10^{-13}\mbox{cm}.\label{scale}
\ee
Kaluza-Klein tower such built is shown in Table 1 where the
comparison with experimentally observed mass spectrum of three-pion
system is also presented.

\begin{table}[tbp]
\begin{center}
\caption{Kaluza-Klein tower of KK excitations of three-pion system
and experimental data.}
\vspace{5mm}
{\large
\begin{tabular}{|c|c|c|c|c|c|}\hline   
 n & $ M_n^{3\pi^0}MeV $ & $ M_n^{\pi^{\pm}2\pi^0}MeV $ & $
 M_n^{\pi^0 2\pi^\pm}MeV $ & $ M_n^{3\pi^\pm}MeV $ & $
 M_{exp}^{3\pi}\,MeV $  \\
 \hline  
1  & 423.62  & 428.02  & 432.42  & 436.81  &  \\
2  & 475.30  & 479.23  & 483.17  & 487.10  &  \\
3  & 550.77  & 554.17  & 557.57  & 560.98  & $\eta(0^{-+})[547]$ \\ 
4  & 641.68  & 644.60  & 647.53  & 650.46  &  \\
5  & 742.38  & 744.91  & 747.44  & 749.98  &  \\
6  & 849.40  & 851.61  & 853.83  & 856.05  &  \\
7  & 960.62  & 962.58  & 964.55  & 966.51  & $\eta'(0^{-+})[958]$ \\
8  & 1074.75 & 1076.51 & 1078.26 & 1080.02 &  \\
9  & 1190.95 & 1192.53 & 1194.12 & 1195.70 & 1194 $\pm$ 14 \\
10 & 1308.66 & 1310.10 & 1311.55 & 1312.99 & 1311.3$\pm$1.6 \\
11 & 1427.51 & 1428.84 & 1430.16 & 1431.49 & 1419 $\pm$ 31 \\
12 & 1547.25 & 1548.47 & 1549.69 & 1550.91 &  \\
13 & 1667.68 & 1668.81 & 1669.94 & 1671.08 & 1667 $\pm$ 4 \\
14 & 1788.65 & 1789.71 & 1790.76 & 1791.82 & 1801 $\pm$ 13 \\
15 & 1910.07 & 1911.06 & 1912.05 & 1913.04 &               \\
16 & 2031.86 & 2032.79 & 2033.72 & 2034.65 & 2030 $\pm$ 50 \\
17 & 2153.95 & 2154.83 & 2155.70 & 2156.58 & 2090 $\pm$ 30 \\
18 & 2276.29 & 2277.12 & 2277.95 & 2278.78 &  \\
19 & 2398.85 & 2399.64 & 2400.43 & 2401.22 &  \\
20 & 2521.69 & 2522.35 & 2523.10 & 2523.85 &  \\
21 & 2644.50 & 2645.22 & 2645.93 & 2646.65 &  \\
22 & 2767.54 & 2768.23 & 2768.91 & 2769.59 &  \\
23 & 2890.71 & 2891.36 & 2892.02 & 2892.67 &  \\
24 & 3013.97 & 3014.60 & 3015.23 & 3015.86 &  \\
25 & 3137.33 & 3137.94 & 3138.54 & 3139.14 &  \\
26 & 3260.78 & 3261.36 & 3261.94 & 3262.52 &  \\
27 & 3384.29 & 3384.85 & 3385.41 & 3385.97 &  \\
28 & 3507.87 & 3508.41 & 3508.95 & 3509.49 &  \\
29 & 3631.51 & 3632.03 & 3632.55 & 3633.08 &  \\
30 & 3755.21 & 3755.71 & 3756.21 & 3756.72 &  \\ \hline
\end{tabular}}
\end{center}
\end{table} 

\begin{table}[hbt]
\begin{center}
\caption{$M_{9}(1191-1196)$--Storey.}
\vspace{5mm}
\begin{tabular}{|c|c|c|c|c|}\hline   
$R(I^GJ^{PC})$ & $ M_R \, MeV $ & $ \Gamma_R \, MeV $ & Reaction &
Collab. \\ \hline   
$h_1(0^-1^{+-})$ & 1190 $\pm$ 60 & 320 $\pm$ 50 & $ \pi p
\rightarrow 3\pi n $ & SPEC 81 \\ 
$\pi(1^-0^{-+})$ & 1190 $\pm$ 30 & 440 $\pm$ 80 & $ \pi^+ Z
\rightarrow Z 3\pi $ & SPEC 84 \\ 
$a_1(1^-1^{++})$ & 1194 $\pm$ 14 & 462 $\pm$ 56 & $ \tau^+
\rightarrow \pi^+\pi^+\pi^-\nu $ & MRK2 86 \\
 & 1208 $\pm$ 15 & 430 $\pm$ 50 & $ pp
\rightarrow pp\pi^+\pi^-\pi^0 $ & OMEG 90 \\
\hline
\end{tabular}
\end{center}
\end{table}

\begin{table}[hbt]
\begin{center}
\caption{$M_{10}(1309-1313)$--Storey.}
\vspace{5mm}
\begin{tabular}{|c|c|c|c|r|}\hline   
$R(I^GJ^{PC})$ & $ M_R \, MeV $ & $ \Gamma_R \, MeV $ & Reaction &
Collab.\ \ \ \\ \hline   
$\pi(1^-0^{-+})$ & 1342 $\pm$ 20 & 220 $\pm$ 70 & $ \pi^- p
\rightarrow p 3\pi $ & OMEG 81 \\
$a_1(1^{-}1^{++})$ & 1280 $\pm$ 30 & 300 $\pm$ 50 & $ \pi^- p
\rightarrow p3\pi$ & CNTR 81 \\
 & 1285 -- 1331 & 619 -- 814 & $ \tau^- \rightarrow
\nu_\tau [3\pi]^-$ & CLEO 99 \\
$a_2(1^{-}2^{++})$ & 1317 $\pm$ 3 & 120 $\pm$ 10 & $ pp \rightarrow
pp\pi^+\pi^-\pi^0$ & WA102 98 \\
 & 1311.3$\pm$1.6 & 103.0$\pm$6.0 & $ \pi^- p
 \rightarrow \pi^+\pi^-\pi^0 n$ & VES 96 \\
 & 1310 $\pm$ 5 & 120 $\pm$ 10 & $ pp \rightarrow pp\pi^+\pi^-\pi^0$
 & OMEG 90 \\
 & 1317 $\pm$ 2 & 96 $\pm$ 9 & $ \pi^- p \rightarrow 3\pi p$ & SPEC
 80 \\
 & 1318 $\pm$ 7 & 112 $\pm$ 18 & $ \pi^+ n \rightarrow p(3\pi)^0 $ &
 DBC 75 \\
 & 1305 $\pm$ 14 & 120 $\pm$ 40 & $ \gamma p \rightarrow
 \eta\pi^+\pi^+\pi^- $ & SHF 93 \\
 & 1310 $\pm$ 2 & 97 $\pm$ 5 & $ \pi^- p \rightarrow 3\pi p$ & OMEG
 81 \\
 & 1306 $\pm$ 4 & 79 $\pm$ 12 & $ \pi^+ p \rightarrow 3\pi p$ & HBC
 70 \\ \hline
\end{tabular}
\end{center}
\end{table}

\begin{table}[hbt]
\begin{center}
\caption{$M_{11}(1428-1432)$--Storey.}
\vspace{5mm}
\begin{tabular}{|c|c|c|c|r|}\hline   
$R(I^GJ^{PC})$ & $ M_R \, MeV $ & $ \Gamma_R \, MeV $ & Reaction &
Collab.\ \  \\ \hline   
$\omega(0^-1^{--})$ & $1400^{\,+100}_{\,-200}$ & 187 $\pm$ 15 & $
e^{+}e^-
\rightarrow \pi^+\pi^-\pi^0 $ & RVUE 98 \\
 & 1419 $\pm$ 31 & 174 $\pm$ 59 & $ e^{+}e^-
\rightarrow \rho\pi $ & DM2 92 \\ \hline
\end{tabular}
\end{center}
\end{table}

\begin{table}[hbt]
\begin{center}
\caption{$M_{13}(1668-1671)$--Storey.}
\vspace{5mm}
\begin{tabular}{|c|c|c|c|r|}\hline   
$R(I^GJ^{PC})$ & $ M_R \, MeV $ & $ \Gamma_R \, MeV $ & Reaction &
Collab.\ \ \\ \hline   
$\omega(0^-1^{--})$ & 1670 $\pm$ 20 & 160 $\pm$ 20 & $\gamma p
\rightarrow
3\pi X$ & OMEG 83 \\
 & 1679 $\pm$ 34 & 99 $\pm$ 49 & $e^+e^- \rightarrow 3\pi$
 & FRAM 80 \\
 & 1652 $\pm$ 17 & 42 $\pm$ 17 & $e^+e^- \rightarrow 3\pi$ &
 OSPK 79 \\
$\omega_{3}(0^-3^{--})$  & 1665.3$\pm$5.2 & 149$\pm$19
& $ \pi^- p \rightarrow \pi^+\pi^-\pi^0 n$ & VES 96 \\
 & 1673 $\pm$ 12 & 173 $\pm$ 16 & $ \pi^+ p \rightarrow \Delta 3\pi$
 & HBC 78 \\
 & 1650 $\pm$ 12 & 253 $\pm$ 39 & $ \pi^- p \rightarrow N3\pi$ & OMEG
 78 \\
 & 1669 $\pm$ 11 & 173 $\pm$ 28 & $ \pi^+ p \rightarrow \Delta^{++}
 3\pi$ & HBC 75 \\
 & 1678 $\pm$ 14 & 167 $\pm$ 40 & $ \pi^+ n \rightarrow  p3\pi^0$ &
 DBC 74 \\
 & 1679 $\pm$ 13 & 155 $\pm$ 40 & $ \pi^+ n \rightarrow  p3\pi^0$ &
 DBC 71 \\
 & 1670 $\pm$ 20 & 100 $\pm$ 40 & $ \pi^+ n \rightarrow  p3\pi^0$ &
 DBC 69 \\
 & 1667 $\pm$ 4 & 168 $\pm$ 10 & AVERAGE & PDG 00 \\
$\pi_{2}(1^-2^{-+})$ & 1676 $\pm$ 6 & 260 $\pm$ 20 & $ \pi^- p
\rightarrow 3\pi p$ & OMEG 81 \\
 & 1657 $\pm$ 14 & 219 $\pm$ 20 & $ \pi p
\rightarrow 3\pi X $ & SPEC 80 \\
 & 1662 $\pm$ 10 & 285 $\pm$ 60 & $ \pi^+ p \rightarrow p 3\pi$ & HBC
 77 \\
 & 1672 $\pm$ 3.5 & 259 $\pm$ 11 & AVERAGE & PDG 00 \\
 \hline
\end{tabular}
\end{center}
\end{table}

\begin{table}[hbt]
\begin{center}
\caption{$M_{14}(1789-1792)$--Storey.} 
\vspace{5mm}
\begin{tabular}{|c|c|c|c|r|}\hline   
$R(I^GJ^{PC})$ & $ M_R \, MeV $ & $ \Gamma_R \, MeV $ & Reaction &
Collab.\ \ \\ \hline 
$a_2(1^-2^{++})$ & 1752$\pm$21$\pm$4 & 150$\pm$110$\pm$34 & $
\gamma\gamma \rightarrow
\pi^+\pi^-\pi^0$ & L3 97 \\
$X(1^-?^{-+})$ & 1763$\pm$20 & 192$\pm$60 & $ \gamma p \rightarrow
n\pi^+\pi^+\pi^-$ & SHF 91 \\
 & 1787$\pm$18 & 118$\pm$60 & $ \gamma p \rightarrow
(p\pi^+)(\pi^+\pi^-\pi^-)$ & SHF 91 \\
 & 1776$\pm$13 & 155$\pm$40 & AVERAGE & PDG 00 \\
$\pi(1^-0^{-+})$ & 1775$\pm$7 & 190$\pm$15 & $ \pi^- A
\rightarrow \pi^+\pi^-\pi^- A$ & VES 95 \\
 & 1770 $\pm$ 30 & 310 $\pm$ 50 & $\pi^- A
\rightarrow 3\pi A$ & SPEC 82 \\
 & 1801 $\pm$ 13 & 210 $\pm$ 15 & AVERAGE & PDG 00 \\
 \hline
\end{tabular}
\end{center}
\end{table}

\begin{table}[hbt]
\begin{center}
\caption{$M_{16}(2032-2035)$--Storey.} 
\vspace{5mm}
\begin{tabular}{|c|c|c|c|r|}\hline   
$R(I^GJ^{PC})$ & $ M_R \, MeV $ & $ \Gamma_R \, MeV $ & Reaction &
Collab.\ \ \\ \hline   
$a_4(1^-4^{++})$ & 2030 $\pm$ 50 & 510 $\pm$ 20 & $\pi^- p
\rightarrow 3\pi n$ & OMEG 78 \\ \hline
\end{tabular}
\end{center}
\end{table}

\begin{table}[tbp]
\begin{center}
\caption{$M_{17}(2154-2157)$--Storey.} 
\vspace{5mm}
\begin{tabular}{|c|c|c|c|r|}\hline   
$R(I^GJ^{PC})$ & $ M_R \, MeV $ & $ \Gamma_R \, MeV $ & Reaction &
Collab.\ \ \\ \hline   
$\pi_2(1^-2^{-+})$ & 2090 $\pm$ 30 & 520 $\pm$ 100 & $\pi^{-}A
\rightarrow \pi^+\pi^-\pi^{-}A$ & VES 95 \\
 & 2100 $\pm$ 150 & 651 $\pm$ 50  & $\pi^{-}p
\rightarrow 3\pi X$  &  CNTR 81 \\ \hline
\end{tabular}
\end{center}
\end{table}
We have used Review of Particle Physics \cite{4} where the
experimental data on mass spectrum of the resonance states of
three-pion system have been extracted from. Again we see from Table 1
that there is a quite remarkable correspondence of the calculated KK
excitations for three-pion system with the experimentally observed
mass spectrum of three-pion resonance states, which we consider as an
additional strong evidence of Kaluza-Klein picture of the world. 

Some known experimental information concerning the experimentally
observed mass spectrum of three-pion system is collected in separate
tables: Table 2 -- Table 8. Certainly, here we have a much more poor
experimental data set compared to the case of two-pion system.
Nevertheless we can learn from these tables a few remarkable facts as
well. 

First of all, as it was mentioned in our previous paper \cite{3},
many different three-pion resonances with the different quantum
numbers may occupy one and the same storey in KK tower. This is a
peculiarity of the systematics provided by Kaluza-Klein picture. 

It is especially pleased for us to emphasize that the experimental
measurement of the $a_2$ meson mass made by Protvino VES
Collaboration \cite{5} with the best world precision 
\[
M(a_2)=1311.3 \pm 1.6(stat) \pm 3.0(syst)\,MeV
\]
is in excellent agreement with the theoretically calculated value  
\[
M_{10}^{\pi^+\pi^-\pi^0}=1311.55\,MeV.
\]
The same is true for the
$\omega_3$ meson where the theoretically calculated mass of KK
excitation in $3\pi^0$ system $M_{13}^{3\pi^0}=1667.68\,MeV$ is in a
very good agreement with PDG AVERAGE value $M(\omega_3)=1667 \pm
4\,MeV$. Moreover, it is very interesting to point out that
theoretical calculation of KK excitations in $\rho\pi$ system by the
formula 
\be
M_n^{\rho\pi} = \sqrt{m_{\rho}^2+\frac{n^2}{R^2}} +
\sqrt{m_{\pi}^2+\frac{n^2}{R^2}},\quad
(n=1,2,3,\ldots),\label{rhopi}
\ee
where we use $m_{\rho}=769.3\,MeV$ for the $\rho$ meson mass from
\cite{4}, gives $M_{10}^{\rho\pi^0}=1310.28\,MeV$ and
$M_{10}^{\rho\pi^{\pm}}=1311.67\,MeV$ which accurately agree with the
experimental measurement of the $a_2$ meson mass provided by VES
Collaboration. This means that $a_2$ meson may manifest itself as a
configuration of $\rho\pi$ system in the main, and this is a very
nontrivial fact. For example, that statement is not true for the
$\omega_3$ meson.

Of course, it would be very desirable to state new experiments to
search a further justification of the systematics provided by Kaluza
and Klein picture of the world, e.g. to fill the empty cells in Table
1. We believe that  this is a quite promising subject of the
investigations in particle and nuclear physics.


\begin{thebibliography}{**}
\bibitem{1}
A.A.~Arkhipov, hep-ph/0208215 (2002); preprint IHEP 2002-43,
Protvino, 2002,
available at http://dbserv.ihep.su/\~{}pubs/prep2002/ps/2002-43.pdf
\bibitem{2}
A.A.~Arkhipov, hep-ph/0302164 (2003).
\bibitem{3}
A.A.~Arkhipov, hep-ph/0302213 (2003).
\bibitem{4}
D.E.~Groom et al., {\it Review of Particle Physics}, Eur. Phys. J.
C{\bf 15}, 401-487 (2000).
\bibitem{5}
D.V.~Amelin et al., Zeit. Phys. C{\bf 70}, 71 (1996).
\end{thebibliography}
\end{document}